\documentstyle[prb,aps,epsf]{revtex}
\begin{document}
%\advance\textheight by 0.3in
%\advance\topmargin by -0.25in
\draft

\twocolumn[\hsize\textwidth\columnwidth\hsize\csname@twocolumnfalse%
\endcsname

\title{A Numerical Study of the Random Transverse-Field Ising Spin
Chain}

\author{A. P. Young}
\address{Department of Physics, University of California, Santa Cruz, 
CA 95064}

\author{H. Rieger}
\address{Institut f\"ur Theoretische Physik, Universit\"at zu K\"oln,
50937 K\"oln, Germany\\
and\\
HLRZ, Forschungszentrum J\"ulich, 52425 J\"ulich, Germany}

\date{\today}

\maketitle

\begin{abstract}
We study numerically the critical region and the 
disordered phase of the
random transverse-field Ising chain.  By using a
mapping of Lieb, Schultz and Mattis to non-interacting fermions, we 
can obtain a numerically exact solution for rather large system sizes, 
$L \le 128$. Our
results confirm the striking predictions of earlier analytical work and, in
addition, give new results for some probability distributions and
scaling functions.
\end{abstract}

\vskip 0.3 truein
]

\section{Introduction}
It has recently become clear that quantum phase transitions\cite{quantum}
in disordered systems are rather different
from phase transitions driven by thermal
fluctuations. In particular,
Griffiths~\cite{griffiths} showed that
the free energy is a non-analytic function of the
magnetic field in part of the disordered 
phase because of rare regions,
which are more strongly correlated than the average and which
are {\em locally ordered}.
However, in a classical system, this effect is very weak,
all the derivatives being finite\cite{essen}. By contrast, in a quantum
system at zero temperature, these effects are much more pronounced.

One model where these effects can be worked out in detail, and where rare,
strongly coupled regions dominate not only the disordered
phase but also the critical region, is the 
one-dimensional random transverse-field Ising
chain with Hamiltonian
\begin{equation}
{\cal H} = -\sum_{i=1}^L J_i \sigma^z_i \sigma^z_{i+1} -
\sum_{i=1}^L h_i \sigma^x_i \ .
\label{ham}
\end{equation}
Here the $\{\sigma^\alpha_i\}$ are Pauli spin matrices, and the
interactions $J_i$ and transverse fields $h_i$ are both independent
random variables, with distributions $\pi(J)$ and $\rho(h)$ respectively.
The lattice size is $L$, which we take to be even,
and periodic boundary conditions are imposed. 
The ground state of this
model is closely related to the finite-temperature behavior of
a two-dimensional classical Ising model with disorder perfectly
correlated along one direction, which was first studied by McCoy and
Wu\cite{mw}.
Subsequently, the quantum model, Eq.~(\ref{ham}), was studied by Shankar
and Murphy\cite{sm}, and recently, in great detail, by Fisher\cite{dsf}. From a
real space renormalization group analysis, which becomes exact on large
scales, Fisher obtained many new results and considerable physical
insight. The purpose of the present study is to investigate the model in
Eq.~(\ref{ham}) numerically, using a powerful technique\cite{lsm}
which is special
to one-dimensional systems, to verify the surprising predictions of
the earlier work\cite{mw,sm,dsf} and to determine certain distributions
and scaling functions which have not yet been calculated analytically. 

In one-dimension one can perform a gauge transformation to make all the
$J_i$ and $h_i$ positive.
Unless otherwise stated,
the numerical work used the following rectangular distribution:
\begin{eqnarray}
\pi(J) & = & 
\left\{
\begin{array}{ll}
1 & \mbox{for $ 0 < J < 1$} \\
0  & \mbox{otherwise}
\end{array}
\right.
\nonumber \\ 
\rho(h) & = & 
\left\{
\begin{array}{ll}
h_0^{-1} & \mbox{for $ 0 < h < h_0$} \\
0  & \mbox{otherwise.}
\end{array}
\right.
\label{dist}
\end{eqnarray}
The model is therefore characterized by a
single control parameter, $h_0$. As discussed in section II,
the critical point is at $h_0 = 1$ (so the distributions of $h$
and $J$ are then the same) and the deviation from criticality is 
conveniently measured by
the parameter $\delta$ in Eq.~(\ref{delta}), where, for the
distribution in Eq.~(\ref{dist}),
\begin{equation}
\delta = {1 \over 2} \ln h_0 \ .
\label{delta_h0}
\end{equation}

Section II discusses the analytical results obtained previously, and
section III reviews the work of Lieb, Schultz and Mattis~\cite{lsm},
Katsura\cite{katsura}
and Pfeuty\cite{pfeuty}
which relates the Hamiltonian to free fermions, and also explains how this
technique can
be implemented numerically for the random case. In section IV the
numerical results for the distribution of the energy gap are shown, while
section V discusses results for the correlation functions. 
Results for the local susceptibility on smaller sizes,
obtained by the Lanczos method, are discussed in section VI, while 
data for the $q=0$ structure factor, which could be measured in a
scattering experiment, are considered in section VII.
Finally,
in section VIII, we summarize our conclusions and discuss the possible
relevance of the results to models in higher dimensions. 

\section{Analytical results}
In this section we summarize the results obtained earlier by McCoy and
Wu\cite{mw}, Shankar and Murphy\cite{sm} and particularly by
Fisher\cite{dsf}. Defining
\begin{eqnarray}
\Delta_h & = & [\ln h]_{\rm av} \nonumber \\
\Delta_J & = & [\ln J]_{\rm av}
\end{eqnarray}
where $[\ldots]_{\rm av}$ denote an average over disorder, the
critical point occurs when
\begin{equation}
\Delta_h = \Delta_J \ .
\end{equation}
Clearly this is satisfied if the distributions of bonds and fields are
equal, and the criticality of the model then follows from duality. A
convenient measure of the deviation from criticality is given by
\begin{equation}
\delta = { \Delta_h - \Delta_J \over 
[(\ln h)^2]_{\rm av} - \Delta_h^2 + [(\ln J)^2]_{\rm av}  - \Delta_J^2} \ .
\label{delta}
\end{equation}

At a quantum critical point one needs to consider the dynamical critical
exponent, $z$, even when determining static critical phenomena, because
statics and dynamics are coupled. The relation between a characteristic
length scale $l$ and the corresponding time scale $\tau$ is then
$\tau \sim l^z$. For the present model one has, at the critical point,
\begin{equation}
z = \infty \quad (\delta = 0) \ ,
\end{equation}
or, more precisely, the time scale varies as the exponential of the
square root of the corresponding length scale. In addition, the
distribution of local relaxation times is predicted to be very broad.
One of the goals of the present work is to determine the form of the
distribution of a related quantity, the gap to the first excited state.

Moving into the disordered phase, there is still a very broad distribution of
relaxation times because of Griffiths singularities, and one can still,
as a result, define a dynamical exponent but this now varies with
$\delta$, diverging as
\begin{equation}
z = { 1 \over 2 \delta } + C + O(\delta) \ ,
\label{zdiverge}
\end{equation}
for $\delta \to 0$, where $C$ is a non-universal constant.
Moving further away
from the critical point, if one reaches a situation where all the fields
are bigger than all the interactions, then 
Griffiths singularities no longer occur
and the distribution of relaxation times becomes
narrow. Denoting the value of $\delta$ where this happens by $\delta_G$,
Griffiths singularities occur in that part of the disordered phase
where\cite{griff_phase} 
\begin{equation}
0 < \delta < \delta_G  \ .
\end{equation}
Approaching the end of the Griffiths phase, one has
\begin{equation}
\lim_{\delta \to \delta_G^-} z = 0 \ .
\end{equation}
For the distribution used in the numerical calculations,
Eq.~(\ref{dist}),
$\delta_G = \infty$ so Griffiths singularities occur throughout the
disordered phase. In the disordered phase,
the magnetization in the
$z$-direction has a singular piece if a uniform field, $H$, coupling to
$\sigma^z$, is added, namely
\begin{equation}
m_{sing} \sim |H|^{1 \over z} \ ,
\end{equation}
so the linear susceptibility diverges over part of the disordered phase,
a result first found by McCoy and Wu\cite{mw}. 

Next we turn to predictions for the correlation functions
\begin{equation}
C_{ij} = \langle \sigma^z_i \sigma^z_j \rangle \ .
\end{equation}
Again there are very big fluctuations, and, as a result, the average and
typical correlations behave quite differently. The average correlation
function,
\begin{equation}
C_{\rm av}(r) = {1 \over L} \sum_{i=1}^L [ \langle \sigma^z_i \sigma^z_{i+r}
\rangle ]_{\rm av} \ ,
\end{equation}
varies as a power of $r$ at criticality,
\begin{equation}
C_{\rm av}(r) \sim {1 \over r^{2 - \phi} } \quad (\delta = 0) \ ,
\label{cav}
\end{equation}
where
\begin{equation}
\phi = {1 + \sqrt{5} \over 2} = 1.61804\ldots
\end{equation}
is the golden mean, so the power in Eq.~(\ref{cav}) is approximately
0.38. Away from criticality, $C_{\rm av}(r)$ decays exponentially at a rate
given by the {\em true} correlation length, $\xi$, where
\begin{equation}
\xi \approx {l_V \over \delta^\nu} \ ,
\label{truexi}
\end{equation}
with
\begin{equation}
\nu = 2 \ .
\end{equation}
The amplitude of the correlation length, $l_V$, is also known and
given by
\begin{equation}
l_V = { 2 \over \mbox{var }[h] + \mbox{var }[J]} \ .
\end{equation}
For the distribution in Eq.~(\ref{dist}) one has
\begin{equation}
l_V = 1 .
\label{lv}
\end{equation}

Scaling theory predicts that
\begin{equation}
{C_{\rm av}(r; \delta) \over C_{\rm av}(r; \delta=0) } =
\bar{C}_{\rm av}(r / \xi) \ ,
\label{cscale}
\end{equation}
where $\bar{C}_{\rm av}$ is a universal scaling function and $\nu$ is the
true correlation function exponent, predicted\cite{dsf} to equal 2.
Fisher\cite{dsf} has calculated the asymptotic form of the scaling
function in Eq.~(\ref{cscale}) for $ r \gg \xi$ and finds
\begin{equation}
\bar{C}_{\rm av}(x) = D{ e^{-x -4.055 x^{1/3}} \over x^{0.451}}
\quad (x \gg 1 )\ ,
\label{asymp}
\end{equation}
where $D$ is an unknown constant, and 0.451 is the numerical value of
$5/6 - (2-\phi)$. 
% the correlation length amplitude, $l_V$, in Eq.~(\ref{lv}) appropriate
%for the distribution in Eq.~(\ref{dist}). 
%The overall prefactor, $D$, is not known.
%though it might be possible to
%relate it to the coefficient of the divergence
%of $C_{\rm av}(r)$ at criticality\cite{dsf:pc}.

The {\em average} correlation function is, however, dominated by rare pairs of
spins which have a correlation function of order unity, much larger
than the typical value, so it is necessary to consider the
distribution of $\ln C(r)$ to get an idea 
of the {\em typical} behavior.
At the critical point 
\begin{equation}
-\ln C(r) \sim \sqrt{r}  \quad (\delta = 0) \ ,
\label{logcr}
\end{equation}
with the coefficient in Eq.~(\ref{logcr}) having a distribution which is
independent of $r$. A goal of the present study is to investigate this
distribution numerically. In the disordered phase, $-\ln C(r) \propto$
$r$ with a coefficient which is {\em self-averaging} for $r \to \infty$, i.e.
\begin{equation}
-\ln C(r) \approx r/\tilde{\xi}  \ ,
\label{eq23}
\end{equation}
for large $r$, where the typical correlation length,
$\tilde{\xi}$, has the behavior
\begin{equation}
\tilde{\xi} \sim {1 \over \delta^{\tilde{\nu}} } \ ,
\label{xityp}
\end{equation}
with
\begin{equation}
\tilde{\nu} = 1 \ .
\end{equation}

The scaling equation corresponding to Eq.~(\ref{cscale}) but for the average
of the {\em log} of the correlation function is
\begin{equation}
\left[ \ln {C(r; \delta) ]_{\rm av}
\over C(r; \delta=0) } \right]_{\rm av}
= \ln \bar{C}_{typ}(r / \tilde{\xi} ) \ ,
\label{lncscale}
\end{equation}
where  $\bar{C}_{typ}$ is a universal scaling function. From
Eqs.~(\ref{logcr}) and (\ref{eq23}) one has, for $ r \gg \tilde{\xi}$,
\begin{equation}
\ln \bar{C}_{typ}(r / \tilde{\xi}) \approx - r/\tilde{\xi} \ .
\label{large_r}
\end{equation}

For correlations of quantities such as 
the energy, which are local in the fermion operators,
see Eqs.~(\ref{spins}) and (\ref{jw}) of the next section,
Shankar and Murphy\cite{sm} obtained more detailed information. They
calculated not only the exponent for the
typical correlation length in Eq.~(\ref{xityp})
but also the amplitude, finding
\begin{equation}
\tilde{\xi}^{-1} = [\ln h]_{\rm av} - [\ln J]_{\rm av} 
\label{xitypex}
\end{equation}
exactly.
%They also computed moments of the distribution of $\ln C(r)$ for
For $r \gg \tilde{\xi}$ the mean of $\ln C_{en}(r)$ is {\em defined} to be
$-r / \tilde{\xi}$ so
\begin{equation}
[\ln C_{en}(r)]_{\rm av}  \approx -\left\{[\ln h]_{\rm av} - [\ln J]]_{\rm
av} \right\} r \ 
\label{lncav}
\end{equation}
in this limit.
The variance of the distribution is also known\cite{dsf:pc} for $r \gg \xi$:
\begin{equation}
\mbox{var } [\ln C_{en}(r)] \approx \left\{\mbox{var }[\ln h] + \mbox{var }
[\ln J] \right\} r  \ .
\label{varlnc}
\end{equation}
Note that the standard deviation of $\ln C_{en}(r)$
is proportional to $r^{1/2}$ whereas
the mean is proportional to $r$, so $\ln C_{en}(r)$ becomes
self-averaging for $ r \gg \tilde{\xi}$.

Fisher\cite{dsf:pc} has suggested that
Eqs.~(\ref{xitypex})-(\ref{varlnc})
might also be true asymptotically for quantities such
as $\sigma^z$ which are {\em not} local in fermion
operators. 
If this is true, then, for the distribution in Eq.~(\ref{dist}), we have 
\begin{eqnarray}
[ \ln C(r; \delta) ]_{\rm av} & \approx & -2 \delta r
\label{ctilde} \\
\mbox{var } [\ln C(r; \delta) ] & \approx & 2 r
\label{varctilde}
\end{eqnarray}
for $r \gg \tilde{\xi}$.

Note that an important feature of these results is that
the true correlation length (which describes the average
correlation function) has a different exponent from that of the typical
correlation length. 

\section{Mapping to free fermions}
The numerical calculations are enormously simplified by relating the
model in Eq.~(\ref{ham}) to {\em non-interacting} fermions. This
technique was first developed for some related quantum spin chain
problems in a beautiful paper by Lieb, Schultz and
Mattis\cite{lsm} and then applied to the {\em non-random} transverse
field Ising chain by Katsura\cite{katsura} and Pfeuty\cite{pfeuty}. 

The starting point is the Jordan-Wigner transformation, which relates the
spin operators to fermion creation and annihilation operators,
$c^\dagger_i$ and $c_i$, by the following transformation:
\begin{eqnarray}
\sigma^z_i & = & a^{\dagger}_i + a_i \nonumber \\
\sigma^y_i &  = & i( a^{\dagger}_i - a_i) \nonumber \\
\sigma^x_i &  = & 1 - 2 a^\dagger_i a_i =  1 - 2 c^\dagger_i c_i \ ,
\label{spins}
\end{eqnarray}
where
\begin{eqnarray}
a^\dagger_i & = & c^\dagger_i \exp\left[ -i\pi \sum_{j=1}^{i-1} c^\dagger_j c_j
\right] \nonumber \\
a_i & = & \exp\left[ -i\pi \sum_{j=1}^{i-1} c^\dagger_j c_j\right] c_i \ .
\label{jw}
\end{eqnarray}
This works because the Pauli spin matrices anti-commute on the same site
but commute on different sites. The ``string operator'' in the
exponentials in Eq.~(\ref{jw}) is just what is needed to insert an
extra minus sign, converting a commutator to an anti-commutator for
different sites. 
The Hamiltonian can then be written
$$
{\cal H} = \sum_{i=1}^L h_i (c_i^\dagger c_i - c_i c^\dagger_i)
- \sum_{i=1}^{L-1} J_i(c^\dagger_i - c_i)( c^\dagger_{i+1} + c_{i+1}) 
$$
\begin{equation}
+ J_L(c^\dagger_L - c_L)( c^\dagger_{1} + c_{1}) \exp(i\pi {\cal N}) \ ,
\label{hamfermi}
\end{equation}
where
\begin{equation}
{\cal N} = \sum_{i=1}^L c^\dagger_i c_i \ ,
\label{N}
\end{equation}
is the number of fermions.
The last term in Eq.~(\ref{hamfermi}) is different from the other
terms involving the $J_i$
since the string operator in Eq.~(\ref{jw}) acts all the
way round the lattice because of periodic boundary conditions. 
Although the number of fermions is not conserved, the parity of that
number {\em is} conserved, so $ \exp(i\pi {\cal N})$ is a constant of
the motion and has the value 1 or $-1$. Hence, the fermion problem must
have antiperiodic boundary conditions if there is an even number of
fermions and periodic boundary conditions if there is an odd number of
fermions. Note that the fermion Hamiltonian, Eq~(\ref{hamfermi}), is
bi-linear in fermion operators, and so describes {\em free fermions}.

For the non-random model\cite{lsm,katsura,pfeuty}
one solves for the single particle eigenstates of Eq.~(\ref{hamfermi})
by (i) a Fourier transform to operators $c^\dagger_k$ and $c_k$,
where $k$ is the wavevector,
followed by (ii) a Bogoliubov-Valatin transformation in which new fermion
creation operators, $\gamma^\dagger_k$, are formed as a linear combination of
$c^\dagger_k$ and $c_{-k}$ in order to remove the terms in $\cal H$
which do not conserve particle number. In the random case, we proceed in
an analogous way. We define a column vector, $\Psi$, and its hermitian
conjugate row vector $\Psi^\dagger$, each of length $2L$, by
\begin{equation}
\Psi^\dagger = (c^\dagger_1, c^\dagger_2, \ldots, c^\dagger_L, c_1, c_2,
\ldots, c_L) \ .
\label{Psi}
\end{equation}
Note that the $\Psi$ and $\Psi^\dagger$ satisfy the fermion
commutation relations
$$
\Psi^\dagger_i \Psi_j + \Psi_j \Psi^\dagger_i = \delta_{ij} 
$$
\begin{equation}
\Psi^\dagger_i \Psi^\dagger_j + \Psi^\dagger_j \Psi^\dagger_i =
\Psi_i  \Psi_j + \Psi_j  \Psi_i = 0, 
\label{comm-rels}
\end{equation}
irrespective of whether $\Psi_i$ refers to a creation or annihilation
operator. 
For reasons that will become clear below,
the Hamiltonian is written in a symmetrical form,
replacing $c_i c_{i+1}$ by $(c_i c_{i+1} -c_{i+1} c_i)/2$,
and $c^\dagger_i c_{i+1}$ by $(c^\dagger_i c_{i+1} -c_{i+1} c^\dagger_i)/2$
etc.  It can then be written in terms of a real-symmetric
$2L \times 2L$ matrix, $\tilde{H}$, as
\begin{equation}
{\cal H} = \Psi^\dagger \tilde{H} \Psi \,
\end{equation}
where $\tilde{H}$ has the form 
\begin{equation}
\tilde{H} = \left[
\begin{array}{rr}
A & B \\
-B & -A
\end{array}
\right] \ ,
\label{htilde}
\end{equation}
where $A$ and $B$ are $L \times L$ matrices with elements given, for
periodic boundary conditions, by
\begin{eqnarray}
A_{i,i} & = & h_i \nonumber \\
A_{i,i+1} & =  - &J_i/2 \nonumber \\
A_{i+1,i} & =  - &J_i/2 \nonumber \\
B_{i,i+1} & =   &J_i/2 \nonumber \\
B_{i+1,i} & =  - &J_i/2 \ ,
\label{blocks}
\end{eqnarray}
where $i+1$ is replaced by $1$ for $i=L$. 
Note that $A$ is symmetric and $B$ is antisymmetric so $\tilde{H}$ is
indeed symmetric as claimed.
For antiperiodic boundary conditions, one changes the
sign of the terms connecting sites $L$ and 1 in Eq.~(\ref{blocks}).

Next we diagonalize $A$ numerically, using standard
routines\cite{recipes}, to find the single particle eigenstates with eigenvalues
$\epsilon_\mu$, $\mu = 1, 2, \ldots 2 L$ and eigenvectors
$\Phi^\dagger_\mu$ which are linear combinations of the $\Psi^\dagger_i$
with real coefficients.
%We define $\Phi\mu$ to be the
%Hermitian conjugate, i.e. the same
%as $\Phi^\dagger_\mu$ but with the $\Psi$-s and $\Psi^\dagger$-s
%interchanged.
We require that the $\Phi^\dagger_\mu$ have the same commutation
relations as the $\Psi^\dagger_i$, see Eq.~(\ref{comm-rels}), which is satisfied
provided the transformation from the $\Psi_i$ to the $\Phi_\mu$ is
orthogonal, which in turn is guaranteed by the symmetry of $\tilde{H}$
that we enforced above. 
If we interchange the $c^\dagger_i$ with the $c_i$ 
in Eq.~(\ref{Psi}) then $\tilde{H}$ changes sign. Hence
the eigenstates come in pairs, with eigenvectors that are Hermitian
conjugates of each other and eigenvalues which are equal in magnitude
and opposite in sign. We can therefore define $\Phi^\dagger_\mu =
\gamma^\dagger_\mu$ if $\epsilon_\mu > 0$ and $\Phi^\dagger_{\mu^\prime} =
\gamma_\mu$ if $\mu^\prime$ is the state with energy $-\epsilon_\mu$.
The Hamiltonian can then be written just in terms of $L$ (rather than
$2L$) modes as
\begin{equation}
{\cal H} = \sum_{\mu=1}^L \epsilon_\mu (\gamma^\dagger_\mu \gamma_\mu -
\gamma_\mu
\gamma^\dagger_\mu) \ ,
\label{hamdiag}
\end{equation}
where all the $\epsilon_\mu$ are now taken to be positive. From
Eqs.~(\ref{hamfermi}) and (\ref{hamdiag}), one sees that if all
the $J_i$ are zero, then the $\epsilon_\mu$ equal the $h_i$, as expected.
We shall denote by ``quasi-particles'' excitations created by
the $\gamma^\dagger_\mu$,
whereas excitations created by the $c^\dagger_i$ will be
called ``bare particles''.

The many-particle states are obtained by either having or not having a 
quasi-particle
in each of the eigenstates. One has to be careful, though, because,
with periodic boundary conditions, the number of bare particles,
$\cal N$ in Eq.~(\ref{N}), must be odd, while for
states with anti-periodic boundary conditions the number must be even.
Thus, to generate all the many-body states one needs to solve the
fermion problem for
{\em both} periodic and anti-periodic boundary conditions, and
keep only {\em half} the states in each case.

In order to determine which states correspond to the
the ground state and the first excited state
it is useful to consider first the non-random case\cite{lsm,pfeuty}. There
the ground state is in the sector with antiperiodic boundary conditions,
and has no quasiparticles, which corresponds to 
$\cal N$ even as required. Hence the ground state energy is given by
\begin{equation}
E_0 = -\sum_{\mu=1}^L \epsilon_\mu^{ap} \ ,
\label{E0}
\end{equation}
where we indicate that the energies are to be evaluated with
antiferromagnetic boundary conditions. 
The first excited state is in the the sector with periodic boundary
conditions. In the disordered phase, there is one quasi-particle, in the
eigenstate with lowest energy, and this state has
an odd-number of bare particles, as required.
Hence the energy of the first excited state of the pure system
in the disordered phase is given by
\begin{equation}
E_1 = \epsilon_1^{p} -\sum_{\mu=2}^L \epsilon_\mu^{p} \quad (\delta > 0)
\ ,
\label{E1a}
\end{equation}
where we have ordered the energies such that $\epsilon_1$ is the
smallest. At the critical point of the non-random model,
$\epsilon_1$ becomes zero. In the conventional point of view, one then says
that $\epsilon_1$ becomes negative in the ordered phase. From our
perspective of numerical calculations, it is more convenient to define
all the $\epsilon_\mu$ to be positive, which means that we are effectively
interchanging the role of the creation and annihilation operators,
$\gamma^\dagger_1$ and $\gamma_1$. Hence,
in our point of view, there are now no quasi-particles,
but this still corresponds to an odd number of bare particles.
From either point of view, the energy of the first excited state
of the pure system in the
ordered phase is given by
\begin{equation}
E_1 =  -\sum_{\mu=1}^L \epsilon_\mu^{p} \quad (\delta < 0)
\ ,
\label{E1b}
\end{equation}
with all the $\epsilon_\mu^p$ taken to be positive.
Note that in the disordered
phase there is a finite gap, $2 \epsilon_1$, in the thermodynamic
limit, whereas in the ordered phase the gap tends exponentially to zero
as $L \to \infty$. This is the manifestation of broken symmetry.
Note also that we can rephrase the result for $E_1$ of
the pure system by saying that it
is given by Eq.~(\ref{E1a}) if the state with no quasi-particles
has an even number of bare particles and by
Eq.~(\ref{E1b}), if it has an odd number (taking all the $\epsilon_\mu$
to be positive).

For the random problem the picture turns out to be very similar. We find
that the ground state energy is given by Eq.~(\ref{E0}) and the
lowest excited state has energy given either by Eq.~(\ref{E1a}) or
Eq.~(\ref{E1b}), depending on whether the state with no quasi-particles
has an even or an odd number of bare particles\cite{comment}, $\cal N$.
The parity of $\cal N$ is determined from Eq.~(\ref{parity}) below.

We have checked that our the code is correct by comparing results for
$E_0$ and $E_1$ for small sizes obtained from this fermion method
with results obtained for
the original problem, Eq.~(\ref{ham}), using both complete diagonalization
and also the Lanczos method. In all cases the results agreed to within
machine precision. 

We now proceed to the calculation of the correlation functions in the
ground state\cite{lsm}. As discussed above, this is in the sector with
anti-periodic boundary conditions, which will be assumed in the rest of
this section, unless otherwise stated.
Assuming, without loss of generality, that $j > i$,
$C_{ij}$ can be expressed in terms of fermions by
\begin{equation}
C_{ij}
= \langle (c^\dagger_i + c_i)
\exp\left[ -i\pi \sum_{l=i}^{j-1} c^\dagger_l c_l \right]
(c^\dagger_j + c_j) \rangle \,
\end{equation}
where the averages are to be evaluated in the ground state.
Now
\begin{eqnarray}
\exp\left[ -i\pi  c^\dagger_l c_l \right]  & = & 
-(c^\dagger_l - c_l) (c^\dagger_l + c_l) 
\label{exponen} \\
& = & (c^\dagger_l + c_l) (c^\dagger_l - c_l) \ ,
\end{eqnarray}
so defining
\begin{eqnarray}
A_l & = & c^\dagger_l + c_l \nonumber \\
B_l & = & c^\dagger_l - c_l \ ,
\end{eqnarray}
and noting that $A_i^2 = 1$, one has 
\begin{equation}
C_{ij} = \langle B_i \left( A_{i+1} B_{i+1} \ldots A_{j-1} B_{j-1} \right) A_j
\rangle \ .
\end{equation}
This rather complicated looking expression can be evaluated using Wick's
theorem. To see this, note first that 
\begin{eqnarray}
\langle A_i A_j \rangle & = & \langle \delta_{ij} - c^\dagger_j c_i +
c^\dagger_i c_j \rangle \nonumber \\
 & = & \delta_{ij} \,
\end{eqnarray}
(since $c^\dagger_j c_i$ and $c^\dagger_i c_j$ are Hermitian conjugates
of each other and a real diagonal matrix element is being 
evaluated) and similarly
\begin{equation}
\langle B_i B_j \rangle = -\delta_{ij}  \ .
\end{equation}
Hence the only non-zero contractions are $\langle A_j B_i \rangle $ and
$\langle B_i A_j \rangle $, since $\langle B_i B_i \rangle $ and
$\langle A_i A_i \rangle $ never occur. Defining
\begin{equation}
\langle B_i A_j \rangle  = - \langle A_j B_i \rangle  = G_{ij} \ ,
\end{equation}
the correlation function is given by a determinant
\begin{equation}
C_{ij} = 
\left|
\begin{array}{cccc}
G_{i, i+1} & G_{i, i+2} & \cdots & G_{ij} \\
G_{i+1, i+1} & G_{i+1, i+2}  & \cdots &  G_{i+1,j} \\
\vdots & \vdots & \ddots & \vdots \\
G_{j-1, i+1} & G_{j-1, i+2}& \cdots & G_{j-1, j} 
\end{array}
\right| \ ,
\label{det}
\end{equation}
which is of size $j-i$.

$G_{ij}$ can be expressed in
terms of the eigenvectors of the matrix $\tilde{H}$
in Eq.~(\ref{htilde}). Let us write 
\begin{eqnarray}
c^\dagger_i + c_i & = & \sum_\mu \phi_{\mu i} (\gamma^\dagger_\mu + \gamma_\mu)
\nonumber \\
c^\dagger_i - c_i & = & \sum_\mu \psi_{\mu i} (\gamma^\dagger_\mu - \gamma_\mu)
\ ,
\end{eqnarray}
where $\psi$ and $\phi$ can be shown to be orthogonal matrices. It
follows that
\begin{eqnarray}
G_{ij} & = & \langle (c^\dagger_i - c_i) (c^\dagger_j + c_j) \rangle
\nonumber \\
 & = & \sum_\mu \psi_{\mu i} \phi_{\mu j}
 \langle (\gamma^\dagger_i - \gamma_i) (\gamma^\dagger_j + \gamma_j) \rangle
\nonumber \\
 & = & - (\psi^T \phi)_{ij} \ ,
\label{G}
\end{eqnarray}
since $\langle \gamma^\dagger \gamma^\dagger \rangle = \langle
\gamma \gamma \rangle  = 0$ and
there are no quasi-particles in the ground state so
$\langle \gamma^\dagger \gamma \rangle = 0$.

Numerically it is straightforward to compute the $G_{ij}$ from
Eq.~(\ref{G}) and then insert the results into Eq.~(\ref{det}) to determine
the $C_{ij}$ for all $i$ and $j$. 

Finally we note that the parity of the number of bare particles,
$\cal N$, in the state with no quasi-particles can also be obtained,
{\em for either boundary condition}, from the $G_{ij}$ since
\begin{eqnarray}
\langle \exp(i\pi {\cal N}) \rangle & = &
\langle \prod_{i=1}^L B_i A_i \rangle \nonumber \\
& = & \det G \ ,
\label{parity}
\end{eqnarray}
where we assumed that $L$ is even, otherwise, from Eq.~(\ref{exponen}),
there would be an additional minus sign.

\section{Results for the energy gap}

For the pure system, the energy gap,
\begin{equation}
\Delta E = E_1 - E_0 \ ,
\end{equation}
is finite in the disordered phase, and tends to zero exponentially with
the size of the system in the ordered phase. Consider now the random case
in the disordered phase, so  $[\ln h ]_{\rm av} > [\ln J ]_{\rm av}$.
Because of statistical fluctuations, there are finite
regions which are locally ordered, i.e. if one were to average
just over one such region
then the inequality would be the other way round. These regions will
have a very small gap. Hence one expects large
sample to sample fluctuations in the gap, especially for big systems.

\begin{figure}[hbt]
\epsfxsize=\columnwidth\epsfbox{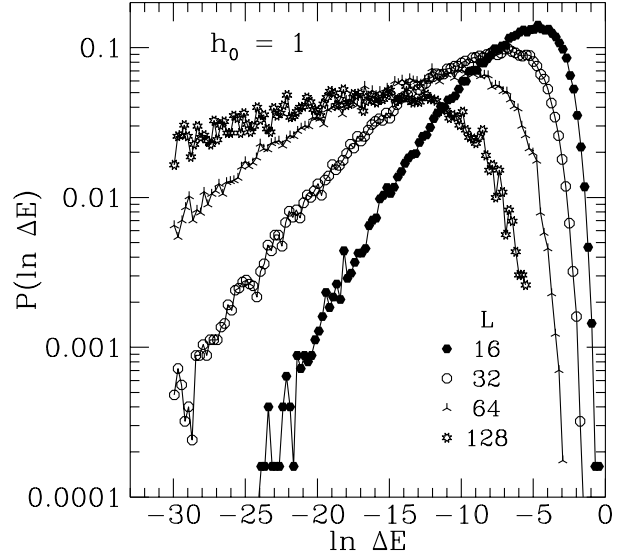}
\caption{A plot of the distribution of the log of the energy gap, $\Delta
E$, at the critical point, $h_0 = 1$,
for lattice sizes between 16 and 128. The distribution was obtained from
the value of the gap for 50000 samples for each size. For the larger sizes,
the distribution is cut off at small values because, in this region,
the gap is essentially zero within double precision accuracy. One sees
that the distribution gets broader and broader as $L$ increases.
}
\label{fig1}
\end{figure}

\begin{figure}[hbt]
\epsfxsize=\columnwidth\epsfbox{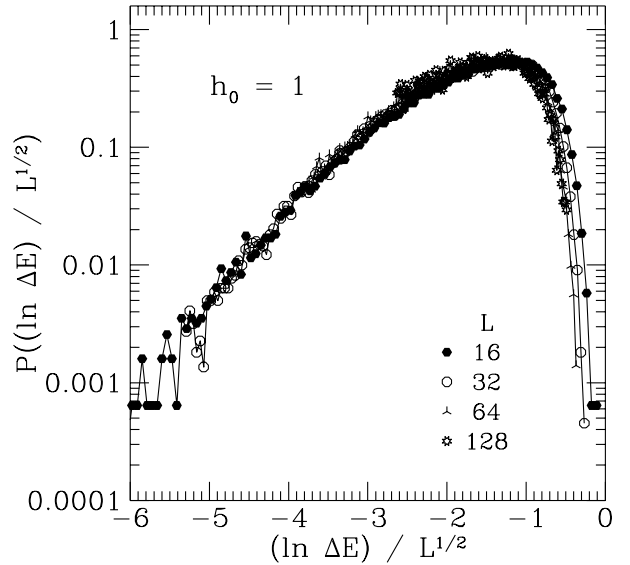}
\caption{A scaling plot of the data in Fig.~\protect\ref{fig1}, assuming that
the log of the energy scale (here the gap) varies as the square root of the
corresponding length scale (here the size of the system).
}
\label{fig2}
\end{figure}

Data for the distribution of $\ln \Delta E$ at
the critical point, $h_0 = 1$, is shown in Fig.~\ref{fig1} for sizes
between 16 and 128.
One sees that the distribution gets broader with
increasing system size. This is clear evidence that $ z =
\infty$ as predicted. The precise prediction is that the log of the
characteristic energy scale should vary as the square root of the
length scale. With this in mind, Fig.~\ref{fig2} shows a scaling
plot for the distribution of $\ln \Delta E / L^{1/2}$, which works quite
well. 

\begin{figure}[hbt]
\epsfxsize=\columnwidth\epsfbox{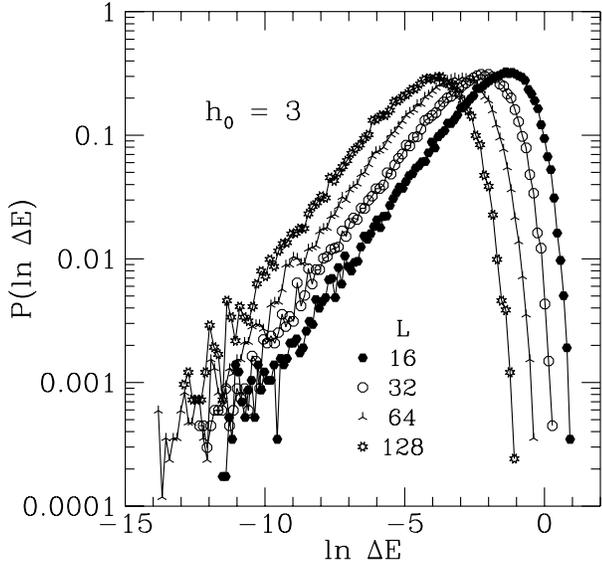}
\caption{A plot of the distribution of the log of the energy gap, $\Delta
E$, in the disordered phase at $h_0 = 3$,
for lattice sizes between 16 and 128. The distribution was obtained from
the value of the gap for 50000 samples for each size. The different curves
are very similar and just shifted relative to each other. 
}
\label{fig3}
\end{figure}

In the disordered phase, the data looks rather different.
Fig.~\ref{fig3} shows the distribution of $\ln \Delta E$, for
$h_0 = 3$. Unlike 
Fig.~\ref{fig1}, the curves for different sizes now look very similar but
shifted horizontally
relative to each other. This implies that the data scales
with a {\em finite} value of $z$, as predicted.

Note that in the region of small gaps, the data in Fig.~\ref{fig3} 
is a straight line indicating a power law distribution of
gaps. This power law behavior is not special to the 1-$d$ problem
discussed here, but is expected quite
generally\cite{th} in the
Griffiths phase for systems with discrete symmetry.
The power is related to $z$ as we shall now see.
Well into the disordered phase, excitations which give a small gap
are well localized so we assume that
the probability of having 
small gap is proportional to the size of the system, $L$.
This assumption is confirmed by the data in 
Fig~\ref{fig3}. Hence, the probability of having a gap between $\Delta E$ and
$\Delta E (1 + \epsilon)$ (for some $\epsilon$) should have the scaling
form, $\epsilon L \Delta E^{1/z}$, so the distribution of gaps,
$P(\Delta E)$, must vary as
\begin{equation}
P(\Delta E) \sim \Delta E^{-1 + 1/z} \ ,
\label{pde}
\end{equation}
in the region of small gaps.
It is tidier to use logarithmic variables, and the corresponding
expression for the distribution of $\ln \Delta E$ is
\begin{equation}
\ln \left[ P(\ln \Delta E) \right] = {1\over z} \ln \Delta E + \mbox{const.} 
\label{plnde}
\end{equation}

\begin{figure}[hbt]
\epsfxsize=\columnwidth\epsfbox{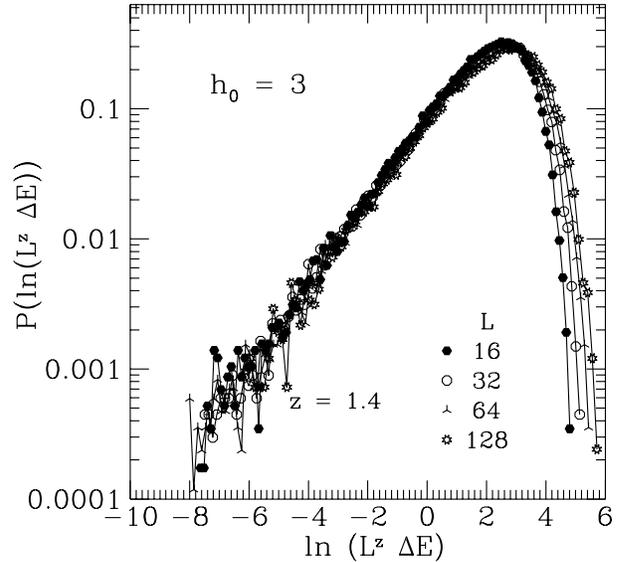}
\caption{A scaling plot of the data in Fig.~\protect\ref{fig3}, assuming 
scaling with a finite value of $z$. The fit here has $z = 1.4$.
}
\label{fig4}
\end{figure}

From the slopes in Fig.~\ref{fig3} we estimate $z \simeq 1.4$, which
gives a satisfactory scaling plot as shown in Fig.~\ref{fig4}. The data
does not collapse so well for large gaps, but this may be outside the
scaling region. 

\begin{figure}[hbt]
\epsfxsize=\columnwidth\epsfbox{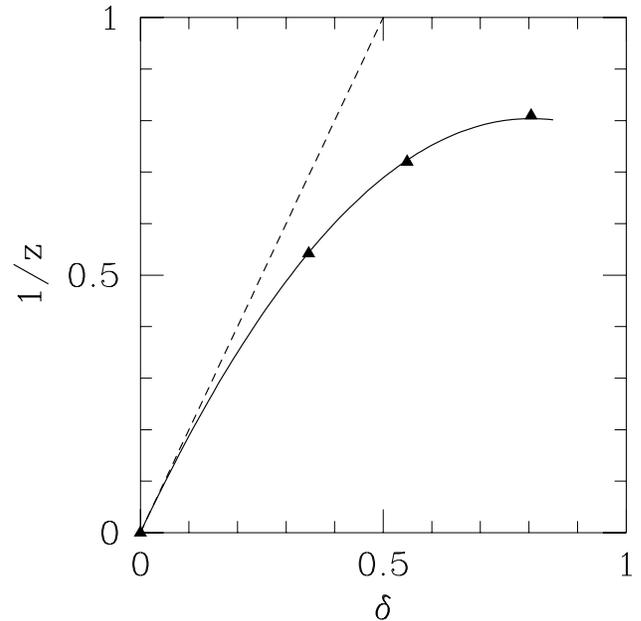}
\caption{Results for $1/z$ against $\delta$, where $z$ is
the dynamical exponent 
and $\delta$ is related to $h_0$ by Eq.~(\protect\ref{delta_h0}).
The solid curve is a fit to $1/z = 2\delta (1 - 2\delta C)$
(which corresponds to the expected form\protect\cite{dsf},
Eq.~(\protect\ref{zdiverge}))
with $C = 0.311$.
The dashed line is $1/z = 2\delta$, the predicted asymptotic form for
$\delta \to 0$.
}
\label{fig5}
\end{figure}

\begin{figure}[hbt]
\epsfxsize=\columnwidth\epsfbox{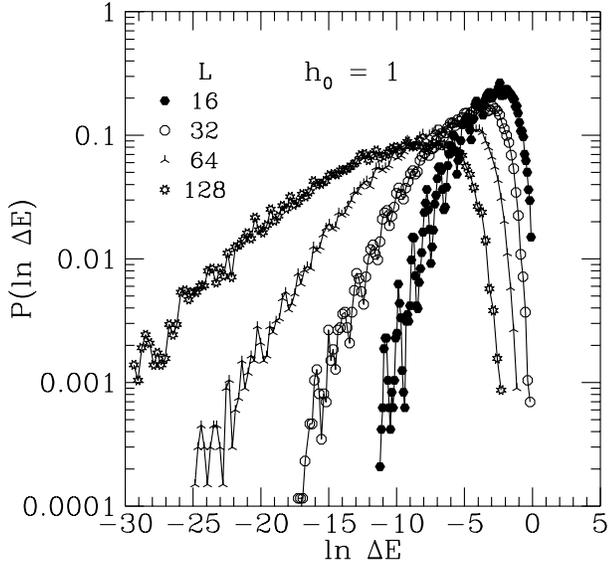}
\caption{A plot of the distribution of the log of the energy gap,
$\Delta E$, at the critical point, $h_0 = 1$, for the bimodal distribution in
Eq.~(\protect\ref{dist2}).
The distribution was obtained from
the value of the gap for 50000 samples for each size.
}
\label{fig6b}
\end{figure}

We have carried out a similar analysis for other values of $h_0$. Close
to the critical point, it is difficult to determine $z$ because the distribution
broadens with increasing size for small sizes
(presumably where $L \le \xi$),
but then the slope of the straight line region starts
to saturate, corresponding to a
large but finite $z$. The sizes that we can study are therefore
in a crossover region between conventional dynamical scaling ($z$
finite) and activated dynamical scaling ($z$ infinite)
so the data does not scale well with any choice of $z$.

Fisher\cite{dsf} has predicted that $z$ is equal to $1/ 2 \delta + C$,
near the critical point, where $C$ is non-universal constant,
see Eq.~(\ref{zdiverge}). We show our estimates for
$1/z$ plotted against $\delta$ (which is related to $h_0$ by
Eq.~(\ref{delta_h0})),
in Fig~\ref{fig5}. Also shown is a fit of $1/z$ to  $2\delta(1- 2\delta
C)$, which corresponds to Eq.~(\ref{zdiverge}) and which works
quite well with $C=0.311$.
%The data does seem to be approaching Eq.~(\ref{zdiverge})
%for small $\delta$. 

\begin{figure}[hbt]
\epsfxsize=\columnwidth\epsfbox{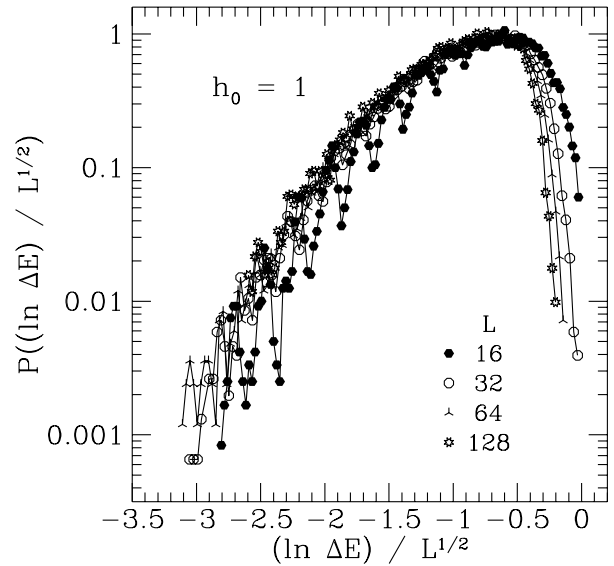}
\caption{A scaling plot of the data in Fig.~\protect\ref{fig6b}.
The collapse of
the data indicates that $z=\infty$, the same as for
the continuous distribution of Eq.~(\protect\ref{dist}),
see Fig.~\protect\ref{fig2}. 
}
\label{fig6}
\end{figure}

The exponents are predicted to be universal, i.e. independent of the 
distributions
$\pi(J)$ and $\rho(h)$ (as long as these don't have anomalously long
tails). To test universality, we also did some calculations for a
bimodal distribution, in which $J$ and $h$ take one of two values,
\begin{eqnarray}
\pi(J) & = & {1\over 2} \left[ \delta(J - 1) + \delta(J - 3) \right]
\nonumber \\ 
\rho(h) & = & {1\over 2} \left[ \delta(h - h_0) + \delta(h - 3 h_0) \right]
\ .
\label{dist2}
\end{eqnarray}
The critical point is at $h_0 = 1$. 
The data for $\Delta E$ at the critical point is shown
in Fig.~\ref{fig6b} and the scaling plot is presented in Fig.~\ref{fig6}.
The data scales reasonably well indicating that
$z=\infty$ at the critical point, just as for the continuous
distribution in Eq.~(\ref{dist}). The data collapse is not as good as
for the continuous distribution, presumably
indicating that the approach to the scaling limit is slower.

\section{Results for correlation functions}
We start by looking at the correlation functions at the critical point
and then discuss our results in the disordered phase. 

\begin{figure}[hbt]
\epsfxsize=\columnwidth\epsfbox{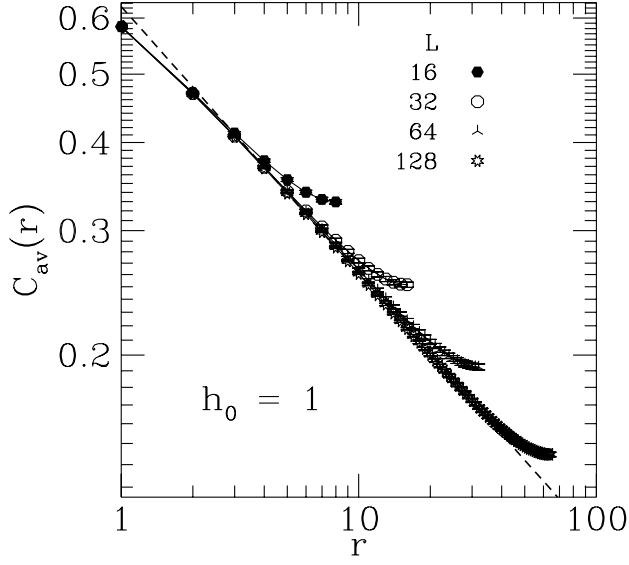}
\caption{
A log-log plot of the average correlation function against distance
at the critical
point. The straight line behavior for larger sizes indicates a power law
variation. The dashed line is a fit to the data for $L=128$ with
$7 \le  r \le 35$ and has slope of $-0.38$ in excellent agreement with
the prediction in Eq.~(\protect\ref{cav}). The results are obtained by
averaging over all pairs of points separated by a distance
$r$ for 10000 samples. 
}
\label{fig7}
\end{figure}

The average correlation function at the critical point is shown in
a log-log plot in Fig~\ref{fig7} for several sizes.
The data for the larger sizes
lie on a straight line, and the dashed line,
which is a fit to the $L=128$ data for $7 \le r \le 35$, has a
slope of $-0.38$, in excellent agreement with Fisher's~\cite{dsf} prediction in
Eq.~(\ref{cav}). 

\begin{figure}[hbt]
\epsfxsize=\columnwidth\epsfbox{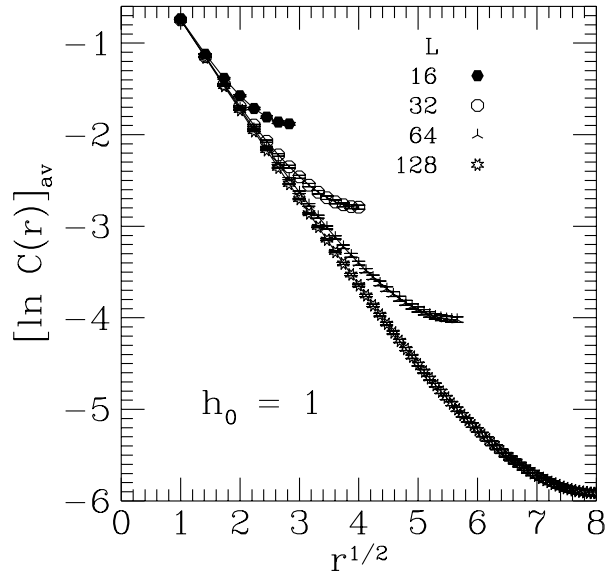}
\caption{
A plot of the average of the log
of the correlation function against $\protect\sqrt{r}$
at the critical
point. The straight line behavior for larger sizes supports the
prediction\protect\cite{sm,dsf} of Eq.~(\protect\ref{logcr}). The
results are obtained by averaging over 10000 samples.
}
\label{fig8}
\end{figure}

A graph of the average of the log of the correlation function (which
corresponds to the log of a {\em typical} correlation function) is shown
in Fig.~\ref{fig8} plotted against $\sqrt{r}$. As expected\cite{sm,dsf} from
Eq.~(\ref{logcr}) the data falls on a straight line. The data in
Figs.~\ref{fig7} and \ref{fig8} indicate that the average and typical
correlation functions do behave very differently at the critical point,
as predicted.

\begin{figure}[hbt]
\epsfxsize=\columnwidth\epsfbox{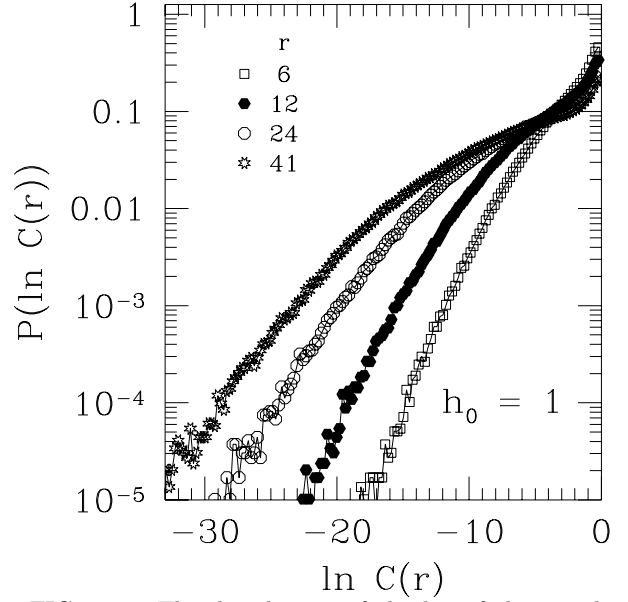}
\caption{
The distribution of the log of the correlation function for different
values of $r$ at the critical point. The data is obtained from 10000
samples of size $L=128$.
}
\label{fig9}
\end{figure}

\begin{figure}[hbt]
\epsfxsize=\columnwidth\epsfbox{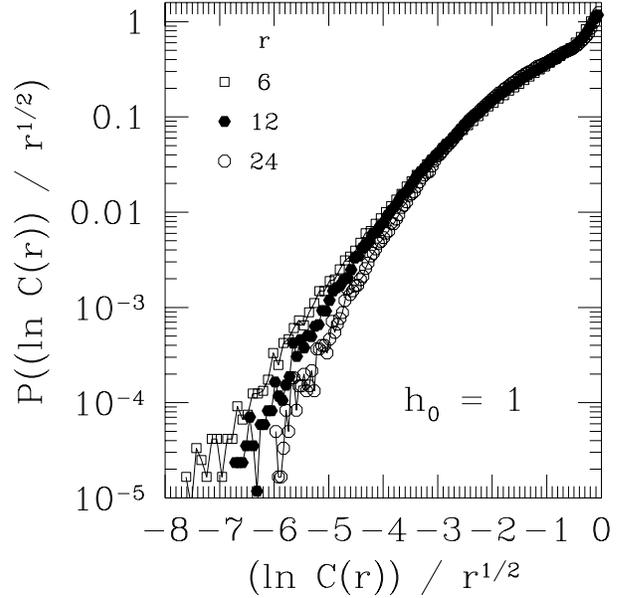}
\caption{
A scaling plot of the data in Fig.~\protect\ref{fig9}
according to the theoretical prediction\protect\cite{dsf}. The
data collapses well in the region of fairly high probability,
including the upturn near the right hand edge (which may indicate a
divergence as the abscissa tends to 0, see the text). 
There are systematic deviations in the tail which may be due to
corrections to scaling.
}
\label{fig10}
\end{figure}

The reason for this difference is that the distribution of $\ln C(r)$ is very
broad, as can be seen in the plot in Fig.~\ref{fig9}. Fisher\cite{dsf} has
predicted that the distribution of $(\ln C(r)) / \sqrt{r}$ should be
universal and independent of $r$ so we show the corresponding scaling plot
in Fig.~\ref{fig10}. Note that the
distribution monotonically decreases as $C(r)$ becomes smaller.
The data scales well for larger values of $C$,
including even the upturn near the
right hand edge of the graph. This is the region where $C(r)$ is
anomalously large and which gives the dominant contribution to
the average correlation function.
An interesting question, then, is whether the value for the average
correlation function is included in
the scaling function for $\ln C(r) / \sqrt{r}$. If so, the the scaling
function will {\em diverge} as a power near the origin\cite{dsf:pc}. To
see this, note that if the probability of having a correlation $C$ at a
distance $r$ only depends on the combination 
\begin{equation}
y = (\ln C) / \sqrt{r}\ ,
\end{equation}
and that if 
\begin{equation}
P(y) \sim {1 \over y^{\lambda}} 
\label{disty}
\end{equation}
for small $y$, then the average value of $C(r)$ is given by
\begin{equation}
[C(r)]_{\rm av} \sim \int_\epsilon^1 dC
\left({\ln C \over r^{1/2} }\right)^{-\lambda} {1
\over r^{1/2}} \ ,
\label{cavlambda}
\end{equation}
assuming that the integral is dominated by the region of small $\ln C$.
Integrating over $C$ gives  a finite number so
\begin{equation}
[C(r)]_{\rm av} \sim {1 \over r^{{1\over 2}(1 - \lambda)}} \ ,
\end{equation}
and comparing with Eq.~(\ref{cav}) yields
\begin{equation}
\lambda = 2 \phi - 3 \simeq 0.24 \ .
\label{lambda}
\end{equation}
\begin{figure}[hbt]
\epsfxsize=\columnwidth\epsfbox{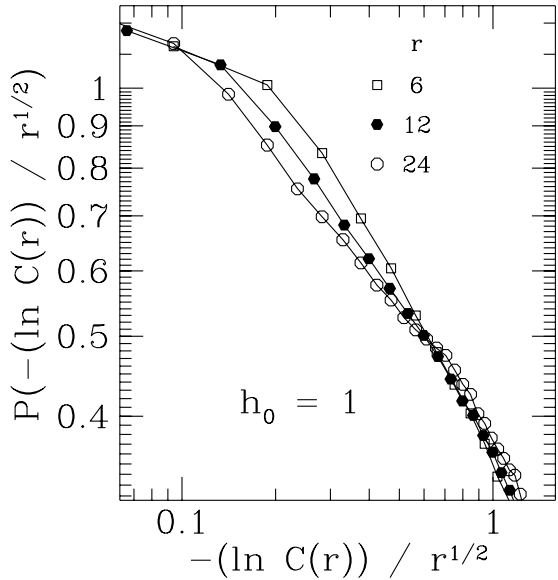}
\caption{
An enlarged log-log plot of the region of the upturn
in Fig.~\protect\ref{fig10} where the abscissa approaches zero..
}
\label{fig10b}
\end{figure}

An enlarged log-log plot of the region of the upturn in Fig.~\ref{fig10}.
is shown in Fig.~\ref{fig10b}. The data does lie on a
rough straight line whose slope decreases (in magnitude)
with increasing $r$. A fit
to the data for $r=24$ in the middle region of the graph has a slope of
about 0.45 (in magnitude), larger than 0.24, but since the effective
slope decreases with increasing $r$ the data does not rule out the
possibility that
the distribution of $(\ln C(r)) / \sqrt{r}$
diverges with an exponent of 0.24 for $r \to \infty$. Even if
the scaling function for  $(\ln C(r)) / \sqrt{r}$ gives the correct {\em power}
for the average correlation function, it does not necessarily mean that
the {\em amplitude} is correct, since there could also be additional
non-universal contributions to the amplitude, outside the
scaling function\cite{dsf:pc}.

We suspect that the systematic deviation in 
the tail of the distribution in Fig.~\ref{fig10} at small values of $C(r)$,
indicates
corrections to scaling for this range of sizes and distances. 

\begin{figure}[hbt]
\epsfxsize=\columnwidth\epsfbox{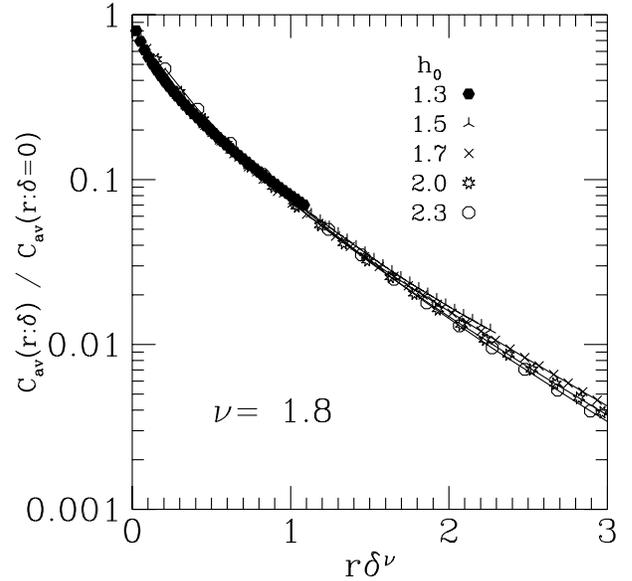}
\caption{
A scaling plot of the average correlation function
in the disordered phase according to
Eq.~(\protect\ref{cscale}). The best fit (shown) is for $\nu = 1.8$,
fairly close to the prediction~\protect\cite{dsf} $\nu = 2$. The data
represents an average over 10000 samples.
}
\label{fig11}
\end{figure}

\begin{figure}[hbt]
\epsfxsize=\columnwidth\epsfbox{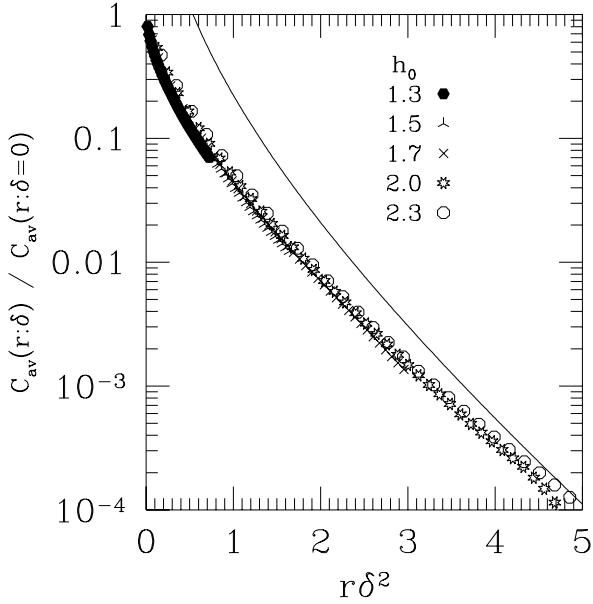}
\caption{
A scaling plot of the average correlation function, according to
Eq.~(\protect\ref{cscale}) with the predicted value
$\nu = 2$. The curve is a plot of
Eq.~(\protect\ref{asymp}) (expected to be
valid for $r \delta^2 \gg 1$) with $D=35$. Note that from
Eqs.~(\protect\ref{truexi}) and (\protect\ref{lv}), one has
$\xi^{-1} = \delta^2$ for the distribution of Eq.~(\protect\ref{dist}).
}
\label{fig11b}
\end{figure}

\begin{figure}[hbt]
\epsfxsize=\columnwidth\epsfbox{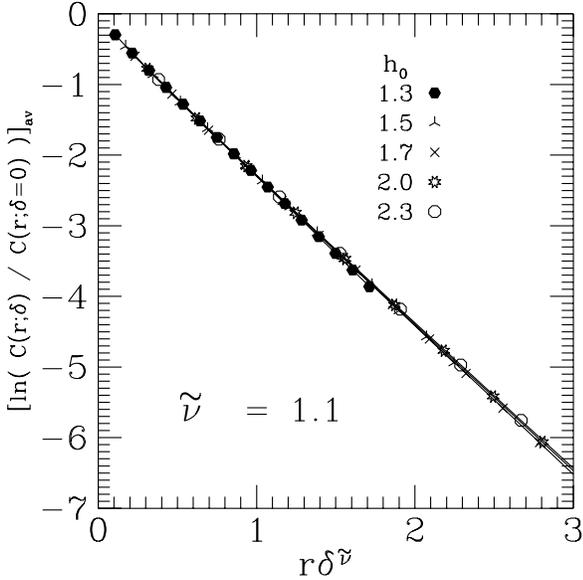}
\caption{
A scaling plot of the average of the log of the correlation function
in the disordered phase according to
Eq.~(\protect\ref{lncscale}). The best fit (shown) is for $\tilde{\nu} = 1.1$,
close to the prediction~\protect\cite{sm,dsf} $\tilde{\nu} = 1$. The data is an
average over 10000 samples. 
}
\label{fig12}
\end{figure}

\begin{figure}[hbt]
\epsfxsize=\columnwidth\epsfbox{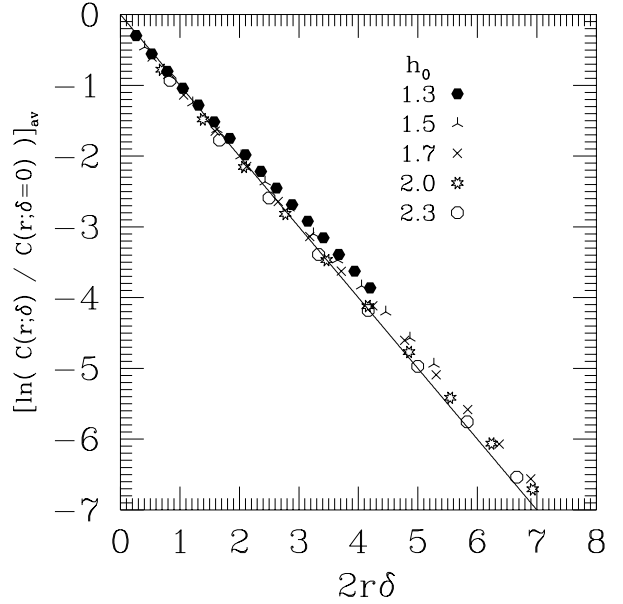}
\caption{
A scaling plot of the average of the log of the correlation function
in the disordered phase according to
Eq.~(\protect\ref{lncscale}), assuming the theoretical value $\tilde{\nu}
= 1$. The solid line is the prediction of Eq.~(\protect\ref{large_r}), which
expected to be valid at large $r$, but also works well down to $r=0$.
It appears that by dividing by the correlation function at criticality,
we incorporate most of the corrections to the asymptotic form in
Eq.~(\protect\ref{large_r}). Note that from Eq.~(\protect\ref{xitypex})
one has $\tilde{\xi}^{-1} = 2 \delta$. 
}
\label{fig12b}
\end{figure}

We now discuss our results for the disordered phase.
The scaling plot corresponding to Eq.~(\ref{cscale}) is shown in
Fig.~\ref{fig11}. The plot has $\nu = 1.8$ which gave
the best fit, and which
agrees fairly well with the
prediction $\nu = 2$. A plot using $\nu = 2$ works
somewhat less well, presumably indicating that there are corrections to scaling 
for this range of lattice sizes and distances.
Fig.~\ref{fig11b} is a scaling plot using the theoretical value $\nu = 2$
which also shows the asymptotic form in Eq.~(\ref{asymp}) with
$D=35$.
Both the data and the prediction of Eq.~(\ref{asymp}), have 
substantial curvature: much more
than in the corresponding data for $\log C(r)$ shown in
Figs.~\ref{fig12} and \ref{fig12b}.
Over the range of accessible values of $r\delta^2$, the data and the
asymptotic prediction do not track each other closely, though it is
possible that they would do so for larger values of $r \delta^2$. 
%Eq.~(\ref{asymp}) tracks the data fairly well
%for large values of the abscissa,
%but, to show convincingly that the data has
%the form in  Eq.~(\ref{asymp}) one would need obtain the value of $D$
%{\em independently}.

The data for the log of the scaling function scales well according to
Eq.~(\ref{lncscale}) , though the best fit has
a slightly different exponent
of 1.1, see Fig.~\ref{fig12}.
Presumably this difference again indicates that there are corrections
to scaling for the sizes and distances studied. Note that the data in
Fig.~\ref{fig12} is {\em close}
to a straight line but there {\em is} statistically
significant curvature. 

Fig.~\ref{fig12b} tests the more stringent prediction\cite{sm,dsf:pc} for
the average of $\ln C(r)$ in the limit $r \gg \tilde{\xi}$
obtained by combining Eq.~(\ref{large_r}) with the assumption that
the expression for
$\tilde{\xi}$ in Eq.~(\ref{xitypex}) is exact for correlations of
$\sigma^z$. One sees that it works quite well.
%We will compare the prediction for the variance of 
%\ln C(r)$ in
%q.~(\ref{varctilde}) with the data in Fig.~\ref{fig13b} below.

\begin{figure}[hbt]
\epsfxsize=\columnwidth\epsfbox{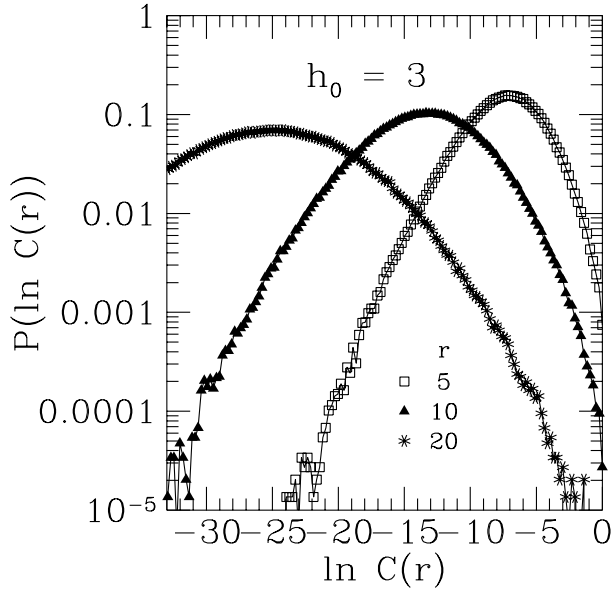}
\caption{
The distribution of the log of the correlation function for different
values of $r$ at $h_0 = 3$ in the disordered phase. The data represents
an average over 10000 samples for $L=64$. Both the most probable value
and the width of the distribution increase with $r$ but the most
probable value increases faster, see Fig.~\protect\ref{fig14}. 
}
\label{fig13}
\end{figure}

\begin{figure}[hbt]
\epsfxsize=\columnwidth\epsfbox{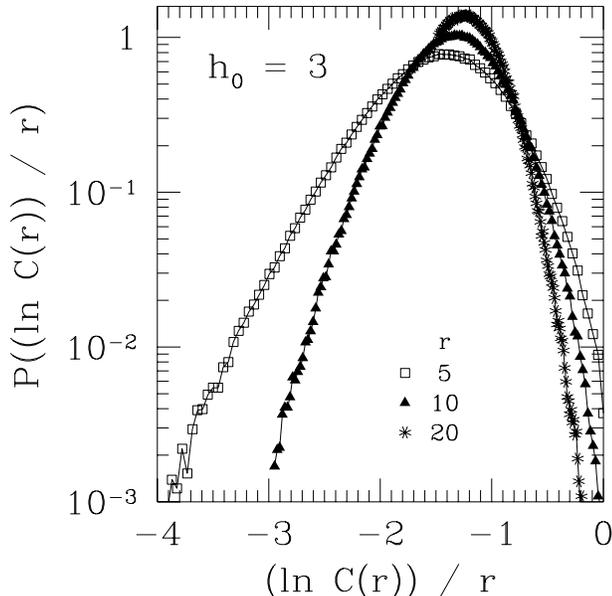}
\caption{
A scaling plot of the data in Fig.~\protect\ref{fig13}. The width of the
distribution is seen to narrow with increasing $r$, as expected. 
}
\label{fig14}
\end{figure}

Although our best estimates of
the critical exponents $\nu$ and $\tilde{\nu}$ do not quite agree
with the theoretical predictions, they are fairly close to those
predictions, and they differ {\em substantially} from
{\em each other}, providing clear evidence that there are
different correlation length
exponents for the average and typical correlation functions. 

\begin{figure}[hbt]
\epsfxsize=\columnwidth\epsfbox{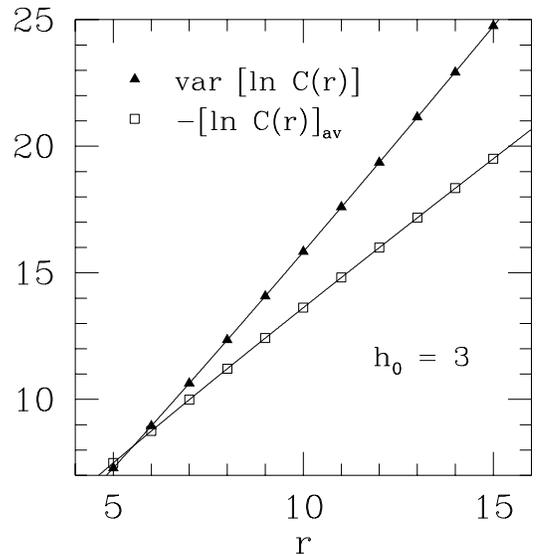}
\caption{
The mean and variance of the distribution of $\ln C(r)$ in the disordered
phase at $h_0=3$ for different values of $r$. Both the mean and variance
are expected be proportional to $r$ at large $r$. The lines are least
squares fits of the form $a + b r^{1/2} + c r$.
This form is motivated by the behavior at criticality, for which the
mean varies as $r^{1/2}$, and, in fact, the $r^{1/2}$ correction to scaling was
effectively removed in Fig.~\protect\ref{fig12b} by factoring out the behavior
at criticality. However it is not clear that the constant and the
$r^{1/2}$ term give all the corrections to scaling in the disordered
phase.
The fit to the mean has
$a = -0.140 , b =1.145 , c = 1.014 $,
while the fit to the variance has $a = 1.025 , b = -1.736, c =2.029 $.
According to the suggestion of Fisher\protect\cite{dsf:pc}, following
Shankar and Murphy\protect\cite{sm}, the leading behavior for large $r$
should be given by Eqs.~(\protect\ref{ctilde}) and (\protect\ref{varctilde}).
Noting that here, $2\delta = \ln 3 \simeq 1.099$,
we see that the large $r$ behavior predicted by the fits is
is in rather good agreement with theory.
}
\label{fig13b}
\end{figure}

Finally, in this section, we look at the {\em distribution} of $\ln C(r)$ for
$r$ larger than either the average or typical correlation lengths. It is
predicted\cite{sm,dsf} that the distribution of $(\ln C(r)) / r$ should
become {\em sharp} at large $r$ in this limit.
Fig.~\ref{fig13} shows data for the
distribution of $\ln C(r)$ at $h_0 = 3$. One sees that both the
peak position and the width increase with increasing $r$, but the
peak position increases faster as can be seen in Fig.~\ref{fig14} which
shows the distribution of $(\ln C(r)) / r$.
In Fig.~\ref{fig13b} we test the more
precise predictions\cite{sm,dsf:pc} for the mean and variance of $\ln C(r)$
given in Eqs.~(\ref{ctilde}) and (\ref{varctilde}), 
for $h_0 = 3$.
The fits give reasonable agreement with
Eqs.~(\ref{ctilde}) and (\ref{varctilde}), but assume a form of
corrections to scaling that we have been unable to justify. 

\section{The local susceptibility}
In this section we discuss the {\em local} susceptibility rather than
the uniform susceptibility, because it has somewhat simpler behavior.
Since it it just involves correlations on a single site, any singularity
must come only from long time-correlations, whereas the uniform
susceptibility involves correlations both in space and time. Our results
for the uniform susceptibility away from the critical point do not scale
in a simple manner, and we suspect that there are logarithmic
corrections, as occurs for bulk behavior at finite
temperature\cite{dsf}. 

%Hence the data is consistent
%with the prediction that the distribution of $(\ln C(r)) / r$ should be
%self averaging for large $r$. 

\begin{figure}[hbt]
\epsfxsize=\columnwidth\epsfbox{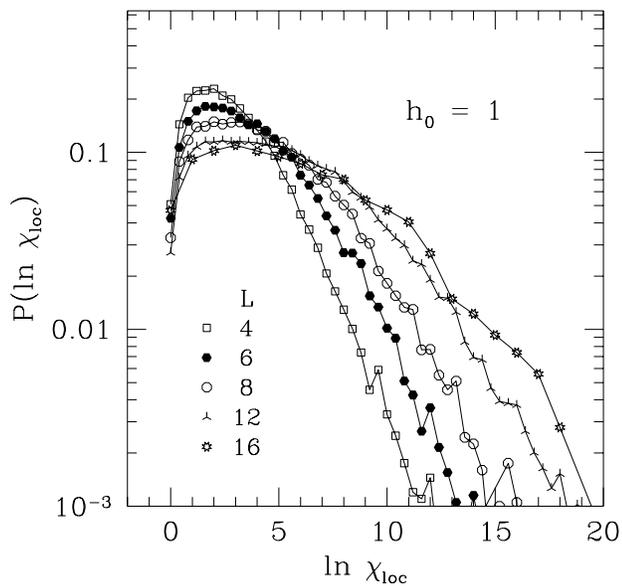}
\caption{
The distribution of the log of the local susceptibility at the critical point
for different sizes obtained by the Lanczos method. The number of
samples is $50000/L$, so that 50000 values for
$\chi_{\rm\scriptscriptstyle loc}$
were obtained for each size. 
}
\label{fig15}
\end{figure}

\begin{figure}[hbt]
\epsfxsize=\columnwidth\epsfbox{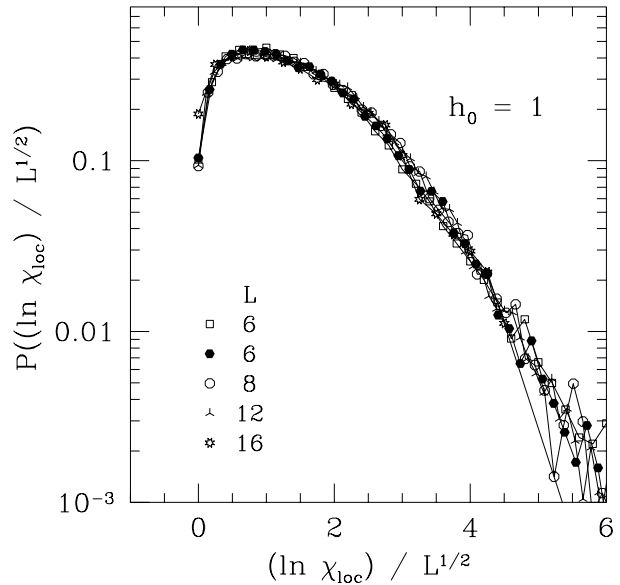}
\caption{
A scaling plot of the data in Fig.~\protect\ref{fig15}. The data scales
in the same way as that for the log of the gap, see Fig.~\protect\ref{fig2}. 
}
\label{fig16}
\end{figure}

Since it is difficult to compute the susceptibility from the fermion
method, particularly with periodic boundary conditions, we have used the
Lanczos diagonalization technique on the original spin Hamiltonian,
Eq.~(\ref{ham}). Of course the price we pay is that the lattices are
much smaller, $L \le 16$. The local susceptibility at $T=0$ is 
given by
\begin{equation}
\chi_{\mbox{\tiny loc}}
= 2 \sum_{n \ne 0} { | \langle 0 | \sigma^z_i | n \rangle |^2
\over E_n - E_0 } \ .
\label{chi}
\end{equation}
where $|n\rangle$ denotes a many body state of the system and
$|0\rangle$ is the ground state.
Because of the form of Eq.~(\ref{chi})
we expect that the scaling of $\chi_{\mbox{\tiny loc}}$ will be
very similar to that of $ 1/\Delta E$. This is indeed the case as 
seen in Fig.~\ref{fig15}, which plots the distribution of
$\ln \chi_{\mbox{\tiny loc}}$
at the critical point. The distribution broadens with system
size, consistent with $z=\infty$.
The data scales in the expected manner, as
shown in Fig.~\ref{fig16}, which is very similar to the corresponding
plot for the energy gap in Fig.~\ref{fig2}.

\begin{figure}[hbt]
\epsfxsize=\columnwidth\epsfbox{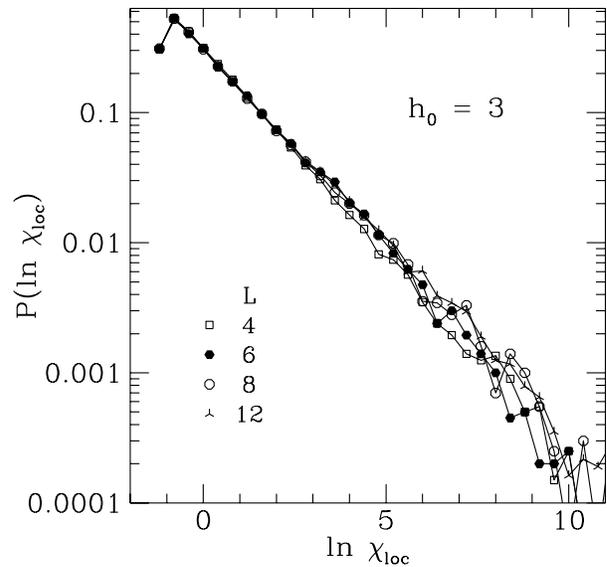}
\caption{
The distribution of the log of the local susceptibility in the disordered
phase for different sizes obtained by the Lanczos method. The number of
samples is 50000 / $L$.
}
\label{fig17}
\end{figure}

Even though the range of sizes used in the Lanczos method is rather
small, it is, nonetheless,
capable of distinguishing $z=\infty$ scaling at the critical
point from finite $z$
scaling $z$ away from the critical point. This can be seen by
comparing Fig.~\ref{fig15} with
Fig.~\ref{fig17}, which plots the
distribution of $\ln \chi_{\mbox{\tiny loc}}$ at $h_0 = 3$.
In Fig.~\ref{fig17} the curves no
longer broaden with increasing $L$ but the distributions are {\em
independent} of size. The reason why there is no size dependence here
but there is in the distributions of $\ln \Delta E$ in Fig.~\ref{fig3}
is easy to understand. For $\Delta E$, we compute the probability {\em per
sample} of getting a certain value, and this is proportional to $L$ in
the disordered phase for small $\Delta E$, since the rare strongly
correlated region can occur anywhere. We used this result in Section
IV to relate the exponent in the distribution to $1/z$, see
Eqs.~(\ref{pde}) and (\ref{plnde}). 
With $\chi_{\mbox{\tiny loc}}$,
however, we compute the probability {\em per site}, so
there is no factor of $L$ and the distribution is independent of size.
This is, of course, the normal state of affairs when the lattice size is much
larger than the correlation length.
The slope of the straight line region in Fig.~\ref{fig17} agrees with
the slopes in Fig.~\ref{fig3} and so gives the same value of z as obtained
from the gap, i.e. $z \simeq 1.4$. 

\section{The Structure Factor}
A scattering experiment measures directly the structure factor, $S( q)$,
defined by
\begin{equation}
S(q) =  {1\over L}\sum_{j,l} C_{jl} e^{i q (j - l)} \ .
\end{equation}

\begin{figure}[hbt]
\epsfxsize=\columnwidth\epsfbox{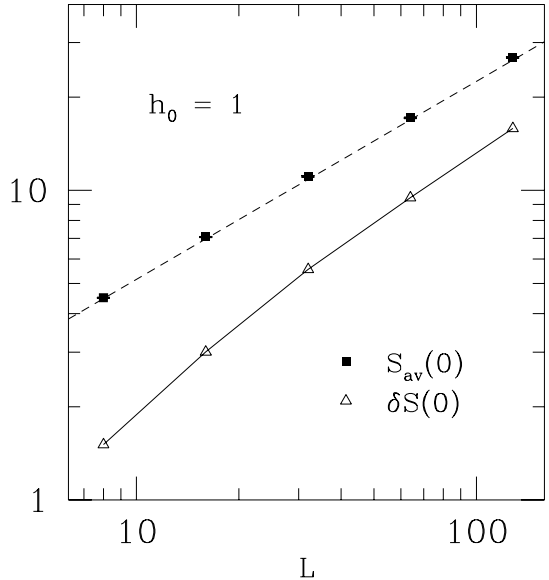}
\caption{
A log-log plot of the
average $q=0$ structure factor, $S_{\rm av}(0)$ and the standard deviation of the
structure factor among different samples, $\delta S(0)$,
for different sizes. The dashed line
is a fit to the data for $S_{\rm av}(0)$ and has slope 0.64, in fairly good
agreement with the theoretical expectation, obtained by integrating
Eq.~(\protect\ref{cav}), of $\phi - 1 \simeq 0.32$. The solid line is
just a guide to the eye. The results are
obtained by averaging over 10000 samples.
}
\label{fig18}
\end{figure}

Although the distribution of individual terms in the sum is broad,
it is interesting, and relevant for experiment, to ask whether there are
large sample to sample fluctuations in the total. We have
attempted to answer this for $q=0$, where fluctuations are expected to
be largest. Fig.~\ref{fig18} shows a log-log plot of
the average of the structure factor, $S_{\rm av}(0)$,
and the standard deviation among different samples, $\delta S(0)$,
plotted against $L$ at the critical point.
From Eq.~(\ref{cav}) one expects the average to vary as $L^{0.62}$ and
the best fit to the numerical data
has a slope of $0.64$, in reasonably good agreement.
One sees that $\delta S(0) < S_{\rm av}(0)$, but the ratio of the width to
the mean stays finite. One expects\cite{dsf:pc} that the distribution of
$S(0) / S_{\rm av}(0)$ will be broad and independent of $L$ at large $L$,
and our data is
consistent with this. Note that although the equal time structure
factor is not self-averaging at $T=0$, its distribution is much less
broad that that of the susceptibility, which involves correlations in
time.
%and, judging from the curvature in the
%data for $\delta S(0)$, it is possible that the ratio $\delta S(0) / S_{\rm
%av}(0)$
%tends to zero in the thermodynamic limit, which would imply
%self averaging, though it is difficult to be sure of this from the data.
%It would be useful
%to understand theoretically how $\delta S(0)$ varies with $L$.
Since the structure factor at the critical point behaves
like the average, rather than the typical correlation
function, we expect that the behavior away from criticality will be
controlled by the average correlation length. 

\section{Conclusions}
We have been able to confirm the many surprising predictions of the
random transverse-Ising spin chain by applying the mapping to free
fermions numerically. In particular we find very broad distributions of
the energy gap and correlation functions, different exponents for the
average and the typical correlation functions, and an infinite value of
the dynamical exponent, $z$, at the critical point. Perhaps the most
interesting {\em new} result is the scaling function for the
distribution of the log of the
correlation function at criticality, shown in Fig.~\ref{fig10}, which
is monotonic and has an upturn as the abscissa approaches zero. If this
indicates the
divergence shown
in Eqs.~(\ref{disty}) and (\ref{lambda}), the scaling function for
$(\ln C(r)) / r^{1/2}$ would also give the
correct exponent (though perhaps not the correct amplitude)
for the average correlation function. 
%and, if so, this part of the
%scaling function then
%determines the average
%correlation function, see Eqs.~(\ref{cavlambda}) and
%(\ref{lambda}). 

We have seen that the width of the distribution of the
the equal time structure factor seems to be comparable to the mean
at $T=0$, though presumably it becomes self-averaging at finite-$T$.
By contrast, the $T=0$ susceptibility and local
susceptibility have enormously broad distributions. One expects that
the susceptibility will also become self averaging
at finite $T$ for sufficiently large $L$, but whether the necessary size
diverges as power law or exponentially as
$T \to 0$ is unclear. We leave this interesting question for future study. 

Crisanti and Rieger\cite{cr} have studied the random transverse Ising chain
by Monte Carlo methods. They took a generalization of the bimodal
distribution in Eq.~(\ref{dist2})
rather than the continuous distribution used here. From the behavior of
various correlation functions they found a finite $z$ at criticality,
which, however, appeared to increase with increasing randomness. We saw
in section IV that corrections to finite-size scaling appear to be larger for
this distribution than for the the continuous one and, furthermore,
%, as noted in the paragraph above,
it is harder to estimate the asymptotic value of $z$ from correlation
functions than from distributions. 
This is presumably why Crisanti and
Rieger\cite{cr} did not find $z=\infty$ in their study.

After this work was largely completed we became aware of related work by
Asakawa and Suzuki\cite{as}, who also used the mapping to free fermions
but used the same distribution as Crisanti and Rieger.
In contrast to our
results, they claim that the exponents depend on the parameters in the
distribution. This is lack of universality is {\em not} predicted by 
theory\cite{sm,dsf} and a possible explanation of the discrepancy is
that not all their data is
in the asymptotic scaling regime, which is likely to be reached for
different lattice sizes for different distributions.

It is interesting to speculate to what extent the results of the
one-dimensional system go over to
higher dimensions. In particular, one would like to know if
$z$ is infinite at the critical point
or takes a finite value for $d > 1$. The results for the local
susceptibility in Section VI indicate that this question {\em can} be answered
even for moderately small lattice sizes {\em provided appropriate
quantities are studied}. The distribution of $\ln \chi_{\mbox{\tiny loc}}$
(or the
log of the gap) seems to be
particularly convenient, since, for finite-$z$, data for different sizes 
look essentially the same, whereas for $z=\infty$ the curves get broader
and broader.
Of course it is still difficult to distinguish a large but
finite $z$ from $z=\infty$, since the two would look the same for small
sizes.
With finite-$z$ scaling, the distribution has power law
behavior, the power being related to $z$ as shown in Eq.~(\ref{plnde}).
It is more difficult to determine $z$ by looking at the decay of
correlation functions, because the asymptotic behavior is only seen at
very large times or distances. Numerical studies in higher dimensions
are likely to use quantum Monte Carlo simulations, because
diagonalization methods, such as Lanczos, can only be carried out on
very small systems and the mapping to free fermions only works in
one-dimension. Unfortunately, there is an
additional difficulty with quantum Monte Carlo, not present here,
because one generally works in imaginary time, which has to be discretized. The
quantum problem is recovered when the number of time slices tends to
infinity, but in practice one can only simulate a finite number. It
is unclear whether the extrapolation to an infinite number of
time slices will pose serious difficulties for the study of Griffiths
singularities and critical phenomena in higher dimensional systems.

We have seen that the disordered Griffiths phase can be conveniently
parameterized by a continuously varying dynamical exponent $z$. This
characterizes the distribution of the energy gap or local susceptibility
for lattice sizes which satisfy the condition, $L \gg \xi$. 
By contrast, at the critical point, the correlation length diverges so
the value of $z$ at criticality involves physics 
in the opposite limit, $L \ll \xi$. It is therefore possible that
the limit of $z(\delta)$ for $\delta \to 0$ is not equal to the value of
$z$ at criticality. 
Both these quantities are
infinite for the transverse field Ising chain, but it would be
interesting to see if there is a difference between them in
higher dimensions. We expect that the results and method of analysis
presented here will provide guidance for such a study.

\acknowledgments
We should like to thank D.~S.~Fisher for many stimulating comments and
suggestions, and for a critical reading of the manuscript.
The work of APY is supported by the National Science
Foundation under grant No. DMR--9411964. 
HR thanks the Physics Department of UCSC for its kind hospitality
and the Deutsche Forschungsgemeinschaft (DFG) for financial support.
His work was performed within the Sonderforschungsbereich SFB 341
K\"oln-Aachen-J\"ulich.


\begin{references}
\bibitem{quantum}
By a quantum phase transition we mean a transition that occurs at $T=0$,
induced by varying some parameter other than the temperature. For the
model in Eq.~(\ref{ham}), this parameter will be the average value of the
transverse field divided by the average interaction, $h_0$ in
Eq.~(\ref{dist}). Alternatively, we can parameterize the model by
$\delta$ defined in Eq.~(\ref{delta}), which is related to $h_0$ by
Eq.~(\ref{delta_h0}).
\bibitem{griff_phase}
We use the term ``Griffiths phase'' to denote the range of values of
$\delta$ where Griffiths
singularities occur at $T=0$. In general, this
covers that part of the disordered phase
where some of the $h_i$ are smaller than some of the $J_i$,
and also that part of
the ordered phase where some of the $J_i$ are smaller than some of the $h_i$.
For the distribution in Eq.~(\ref{dist}), which has no lower bound on the
$h_i$ or $J_i$, the Griffiths phase occurs for {\em all} $\delta$. In
this paper we do not consider the ordered phase ($\delta < 0$).
\bibitem{comment}
For the random problem, as well as for the pure problem,
one eigenvalue hits zero
as $h_0$ (or equivalently $\delta$) is varied.
For the pure system this occurs
precisely at the bulk critical point, but, for the random case, the
point where the eigenvalue vanishes depends on the particular
realization of the disorder, the values for different samples
being scattered around the critical point. 
Hence, for the pure system,
one knows to choose Eq.~(\ref{E1a}) or (\ref{E1b}) for $E_1$
depending on whether
the system is in the disordered or ordered phase, but
there is no such simple
criterion for the random case. We therefore need to explicitly check
whether the number of bare particles, $\cal N$, is even or odd
in the state with no quasi-particles.
\end{references}
\end{document}